\documentclass{article}
\usepackage[utf8]{inputenc}
\usepackage{fullpage}
\usepackage[position=b]{subcaption}
\usepackage{chemformula}

\usepackage{amsmath,amsfonts,amssymb,amsthm}
\usepackage{graphicx}
\usepackage{dsfont}
\usepackage{comment}
\usepackage{url}

\newcommand{\R}{\mathbb R}

\newcommand{\br}{\mathbf r}

\newcommand{\bR}{\mathbf R}

\newcommand{\be}{\begin{equation}}
\newcommand{\ee}{\end{equation}}
\newcommand{\ba}{\begin{eqnarray}}
\newcommand{\ea}{\end{eqnarray}}

\title{Computing photoionization spectra in Gaussian basis sets}
\author{Ivan Duchemin\footnote{Université Grenoble Alpes, CEA, IRIG-MEM-L
    Sim, 38054 Grenoble, France}, Antoine Levitt\footnote{Université Paris-Saclay, CNRS, Laboratoire de mathématiques d’Orsay,
91405 Orsay, France}}
\date{}

\begin{document}

\maketitle

\begin{abstract}
  We present a method to compute the photoionization spectra of atoms
  and molecules in linear response time-dependent density functional
  theory. The electronic orbital variations corresponding to ionized
  electrons are expanded on a basis set of delocalized functions
  obtained as the solution of the inhomogeneous Helmholtz equation
  with gaussian basis set functions as right-hand side. The resulting
  scheme is able to reproduce photoionization spectra without any need
  for artificial regularization or localization. We demonstrate that
  it is able to produce accurate spectra for semilocal
  exchange-correlation functionals even using relatively small
  standard gaussian basis sets.
\end{abstract}

\section{Introduction}
Time-dependent functional theory in the linear response regime
(LR-TDDFT) is widely used to compute excitation energies of molecules.
The traditional way to compute these properties involves solving the
Casida equations \cite{casida1995time}, which are the discretized LR-TDDFT equations in a
basis set of localized functions (typically, gaussian basis sets).
This is very convenient for the excitations lying below the ionization
potential, for which the orbital variations are localized functions.
This however fails to reproduce the photoionization spectrum, turning
a continuous function into a series of infinitely sharp peaks (Dirac
deltas). These peaks approximate the function in a weak sense (when
integrated against smooth functions) in the limit of complete basis
sets \cite{dupuy2022finite}, but pointwise values are not easily obtained.

Various techniques have been used to remedy this. The simplest and
most general technique is to use an artificial dissipation parameter
$\eta$. This effectively adds a constant imaginary part to the
Hamiltonian, broadening the infinitely sharp peaks and producing a
continuous function. Another related technique is the use of complex
absorbing potentials, where the imaginary part is only made to act
away from the molecule, effectively attempting to broaden only the
states that correspond to ionized electrons \cite{muga2004complex}. Another variation on the
same idea is to add the imaginary part to the molecular orbitals
energies directly \cite{coccia2017ab}.
When implemented in the time domain, these techniques can be
understood as adding an artificial dissipation, resulting in states
with finite lifetimes (complex energies).

These schemes shift the poles of the response function away from the real axis
into the lower complex plane, resulting in response functions that can mathematically be
expressed as sums of Lorentzians. Howerver, this does not respect the
mathematical structure of photoionization spectra (infinitely sharp
peaks before the ionization potential, continuous function with sharp
changes around the ionization potentials as well as possible
resonances). As a result, when the basis sets are small (as is the case for typical
gaussian basis sets), spectra are often unsatisfactory.

Another class of methods attempts to recover the continuous
photoionization spectrum from the discrete peaks by fitting rational
functions to quantities that are computable on localized basis sets,
such as moments of the oscillator strengths or values of the
polarizability in the complex plane: examples include Stieljes imaging
or Padé extrapolation (see \cite{tenorio2021molecular} for a review).
These methods are intrinsically numerically unstable and need a
non-trivial manual parameter selection.

More sophisticated schemes like the complex scaling and exterior
complex scaling methods use the analytic continuation of the solutions
to transform oscillatory tails into decaying ones, resulting in more
accurate spectra \cite{cerioni2013accurate}. These schemes are however non-trivial to implement,
and require significant fine-tuning of parameters (as do the
dissipation-based schemes).

A more principled class of methods involve solving an approximate form
of the equation exactly outside of the computational domain, resulting
in a Dirichlet-to-Neumann map that acts as an effective boundary
condition \cite{schwinn2022photoionization}. These schemes are however hard to implement in systems
without spherical symmetry.

In this paper, inspired by a method we recently developed to compute
resonances in locally perturbed periodic media
\cite{duchemin2023efficient}, we propose a new scheme that does not
rely on artificial dissipation or localization methods, and works for
arbitrary molecules without symmetries. The method can be summarized
as follows. The LR-TDDFT equations for the orbital variations
$\delta \psi_{i}^{\pm}$ are of the Helmholtz form
\begin{align*}
  (\pm \omega+i0^{+} + \varepsilon_{i} + \tfrac 1 2 \Delta) \delta\psi_{i}^{\pm} = f
\end{align*}
where $f$ is a localized function (that depends self-consistently on all the
$\delta \psi_{i}^{\pm}$) and $\varepsilon_{i} < 0$ are the Kohn-Sham
one-particle energies. The $i0^{+}$ in this equation means that the
equation is to be solved for finite positive values $i \eta$, and $\eta$ is to
be taken to tend to zero after the calculation.

From the form of the Green function of the operator
$\pm \omega+i0^{+} + \varepsilon_{i} + \tfrac 1 2 \Delta$, it can be
seen that, in the limit $\eta \to 0^{+}$, for
$\omega > - \varepsilon_{i}$, $\delta \psi_{i}^{+}$ will be
delocalized. Physically, $\delta \psi_{i}^{+}$ represents an electron
ionization; the positive values of $\eta$ correspond to imposing an
outgoing wave boundary condition. It is not appropriate to discretize
the delocalized $\delta \psi_{i}^{+}$ on a localized basis set; in
fact, doing so results in a singular photoionization limit that, in
the limit $\eta \to 0^{+}$, tends to a finite sum of Dirac masses
located at the Casida excitation energies. These then have to be
regularized to obtain a continuous spectrum. Instead, we perform the
change of variable
\begin{align*}
  \delta \psi_{i}^{+} = (\omega+i0^{+} + \varepsilon_{i} + \tfrac 1 2 \Delta)^{-1} \delta \phi_{i}^{+},
\end{align*}
and discretize $\phi_{i}^{+}$ (which is localized) on a standard
gaussian basis set. Equivalently, we discretize $\delta \psi_{i}^{+}$ on a basis set
consisting of solutions of the Helmholtz equation with gaussian basis
set functions as right-hand side.

By respecting the structure of the equations
(localized excitation or delocalized ionization, depending on the value of
$\pm \omega + \varepsilon_{i}$), this method allows us to compute
photoionization spectra directly. Compared to the standard method of
solving the Casida equations, it is much more accurate, resulting in
accurate spectra using very moderate basis sets. It can be easily
adapted to compute resonances, and is fully general, being suited to
molecules as well as atoms. The flip side is an added computational
cost, which, although formally having the same cubic scaling as usual
TDDFT methods, has a higher prefactor, mostly due to the need for
frequency-dependent integrals on a real space grid. These are however
highly parallelizable.

\section{Methods}
\subsection{Model}
We consider a molecule with $N$ spin-paired electrons modeled using
time-dependent adiabatic Kohn-Sham density functional theory with a semilocal functional, and
an electrostatic potential $V_{\rm nucl}$ originating from the nuclei.
The ground-state orbitals $\psi_{i}$ satisfy, for all
$i \in \{1, \dots, N/2\}$,
\begin{align*}
  H[\rho] \psi_{i} = \varepsilon_{i} \psi_{i}
\end{align*}
where, in atomic units,
\begin{align*}
  H[\rho] &= - \tfrac 1 2 \Delta + V_{\rm tot}[\rho]\\
  V_{\rm tot} &= V_{\rm nucl} + V_{\rm Hxc}[\rho]\\
  V_{\rm Hxc}[\rho](\br) &= \int\frac{\rho(\br')}{|\br-\br'|} d\br' + v_{\rm xc}(\rho(\br))
\end{align*}
The total density is
$\rho(\br) = 2\sum_{i=1}^{N/2} |\psi_{i}(\br)|^{2}$. We choose the
ground-state orbitals to be real for simplicity, but the scheme
extends trivially to complex orbitals.

Consider now a perturbing potential $\delta V_{\mathcal P}$. In the
time-harmonic regime, the first-order response can be described by the Sternheimer
equations \cite{schwinn2022photoionization}
\begin{align*}
  (\pm \omega+i\eta + \varepsilon_{i} - H[\rho]) \delta\psi_{i}^{\pm} - (f_{\rm HXC} \delta \rho) \psi_{i} = \delta V_{\mathcal P} \psi_{i}
\end{align*}
The variation in the Hartree-exchange-correlation kernel is given by
\begin{align*}
  (f_{\rm HXC} \delta \rho)(\br) = \int\frac{\delta\rho(\br')}{|\br-\br'|} d\br' + v_{\rm xc}'(\rho(\br)) \delta\rho(\br).
\end{align*}
and the variation in density by
\begin{align*}
  \delta \rho(\br) = 2 \sum_{i=1}^{N/2} \psi_{i}(\br) (\delta\psi_{i}^{+}(\br) + \delta \psi_{i}^{-}(\br)).
\end{align*}
This defines the total variation in polarization
\begin{align*}
  {\bf \delta P} = \int \br \delta \rho(\br) d \br.
\end{align*}
When $\delta V_{\mathcal P}(\br) = - {\bf e} \cdot \br$, the linear
relationship ${\bf \delta P} = \alpha(\omega+i\eta) {\bf e}$ defines
the polarizability tensor $\alpha$,
a $3\times 3$ matrix. Finally, the photoionization
cross-section is given by
\begin{align*}
  \sigma(\omega) = \lim_{\eta \to 0^{+}} \frac {4\pi \omega}{c}\frac 1 3{\rm Tr}(\alpha(\omega+i\eta)),
\end{align*}
with $c$ the speed of light, and will be our main observable of interest.

\subsection{Integral form and delocalization}

The standard approach to discretizing these equations is to expand
$\delta \psi_{i}$ in a basis, which leads to the usual Casida
equations. This is however inefficient when $\omega$ is greater than
$-\varepsilon_{i}$, at which point the electron becomes ionized and
$\delta \psi_{i}^{+}$ is delocalized. To see this, we
write the Sternheimer equations as
\begin{align}
  \label{eq:sternheimer}
  (\pm \omega+i\eta + \varepsilon_{i} + \tfrac 1 2 \Delta) \delta\psi_{i}^{\pm} =  V_{\rm tot} \delta \psi_{i}^{\pm} + (f_{\rm HXC} \delta \rho) \psi_{i} + \delta V_{\mathcal P} \psi_{i}
\end{align}
Since $V_{\rm tot}$ and $\psi_{i}$ are localized, the right-hand side is
localized. Introduce the free Green's function $G_{0}(\br,\br';z)$, the kernel
of the inverse of the operator $z + \tfrac 1 2 \Delta$, well-defined
when $\eta > 0$. We can then reformulate the Sternheimer equations in
integral form
\begin{align*}
  \delta \psi_{i}^{\pm} = G_{0}(\pm\omega+i\eta+\varepsilon_{i})\Big(V_{\rm tot} \delta \psi_{i}^{\pm} + (f_{\rm HXC} \delta \rho) \psi_{i} + \delta V_{\mathcal P} \psi_{i}\Big)
\end{align*}
An explicit computation shows that the kernel of the Green function is
given by
\begin{align*}
  G_{0}(\br,\br',z) = - \frac 1 {2\pi} \frac{e^{i k(z) |\br-\br'|}}{|\br-\br'|}
\end{align*}
where $k(z)$ is the square root of $2z$ with positive imaginary part
(so that $G_{0}$ is localized when $\eta > 0$).

When $z = \pm \omega + i \eta + \varepsilon_{i}$ approaches the real axis, this Green function is oscillatory
when ${\rm Re}(z) > 0$, and decaying when ${\rm Re}(z) < 0$. In the region $\omega > 0$,
$-\omega+\varepsilon_{i}$ is always negative, and therefore
$\delta \psi_{i}^{-}$ will be localized. However, the behavior of
$\psi_{i}^{+}$ depends on whether $\omega < -\varepsilon_{i}$ or
$\omega > \varepsilon_{i}$ (below or above ionization threshold). In
the case $\omega < -\varepsilon_{i}$, $\delta \psi_{i}^{+}$ will be
localized; in the case $\omega < -\varepsilon_{i}$, it will be
oscillatory. This explains why usual methods, based on the
discretization of $\delta \psi_{i}$, have trouble reproducing the
ionization region $\omega > -\varepsilon_{i}$.

\subsection{Our method}

Our approach is to use the change of variables
\begin{align*}
  \delta \psi_{i}^{+} = G_{0}(\pm\omega+i\eta+\varepsilon_{i}) \delta\phi_{i}^{+}\; \text{ when $\omega + \varepsilon_{i} > 0$}
\end{align*}
and to discretize $\delta \phi_{i}^{+}$ in a Gaussian basis set
instead of $\delta \psi_{i}^{+}$ directly. This ensures automatically
the correct asymptotic behavior for $\delta \psi_{i}^{+}$. The $\delta
\phi_{i}^{+}$ satisfy the integral equation
\begin{align}
  \label{eq:sternheimer_delta_phi}
  \delta \phi_{i}^{+} - V_{\rm tot} G_{0}(+\omega+i\eta+\varepsilon_{i}) \delta \phi_{i}^{+} - (f_{\rm HXC} \delta \rho) \psi_{i} =  \delta V_{\mathcal P} \psi_{i}
\end{align}

When $\pm \omega + \varepsilon_{i} \le 0$, we discretize
$\delta \psi_{i}^{\pm}$ in the usual way on a basis of Gaussian-type
orbitals $(\chi_{\alpha})_{\alpha=1,\dots,N_{b}}$. When
$\pm \omega + \varepsilon_{i} > 0$, we discretize
$\delta \phi_{i}^{\pm}$ on the basis:
\begin{align*}
  \delta \phi_{i}^{+}(\br) &= \sum_{\alpha=1}^{N_{b}} a_{i\alpha}^{+} \chi_{\alpha}(\br)\; \text{ when $\omega + \varepsilon_{i} > 0$}\\
  \delta \psi_{i}^{\pm}(\br) &= \sum_{\alpha=1}^{N_{b}} b_{i\alpha}^{\pm} \chi_{\alpha}(\br)\; \text{ otherwise}
\end{align*}
This sets up a linear system in the coefficients $(a,b)$.

Note that from \eqref{eq:sternheimer_delta_phi} it follows that
$\delta \phi_{i}^{+}$ is as localized as $V_{\rm tot} \psi_{i}^{+}$.
Therefore, the decay of the total mean-field potential determines the
localization of $\delta \phi_{i}^{+}$, and therefore the effectiveness
of the numerical method. When using semilocal density functionals
(with exponentially decaying exchange-correlation potentials), this is
determined by the electrostatic potential. We can therefore rank
systems by decreasing order of localization: atoms (exponentially
decaying potential), nonpolar molecules (potential decaying as
$1/r^{3}$), polar molecules ($1/r^{2}$), charged systems ($1/r$). When
using hybrid functionals incorporing Hartree-Fock exchange, the effective potential seen by ionized
electrons behaves as $1/r$ \cite{schwinn2022photoionization}, and we
expect our method to have difficulties.
\subsection{Solution of the linear system}
We project the Sternheimer equations \eqref{eq:sternheimer}
and \eqref{eq:sternheimer_delta_phi} on the basis to obtain
\begin{align}
  \label{eq:integral_eq_discretized}
  \begin{cases}
  (S-K_{i}^{+}) a_{i}^{+} - g_{i}[a,b] = h_{i}&\text{ when $\omega + \varepsilon_{i} > 0$}\\
  ((\pm \omega + \varepsilon_{i}+i\eta)S - H) b_{i}^{\pm} - g_{i}[a,b] = h_{i}&\text{ otherwise},
  \end{cases}
\end{align}
where
\begin{align*}
  S_{\alpha\beta} &= \langle  \chi_{\alpha}| \chi_{\beta} \rangle\\
  H_{\alpha\beta} &= \langle  \chi_{\alpha}| - \tfrac 1 2 \Delta + V_{\rm tot} | \chi_{\beta} \rangle\\
  K_{i\alpha\beta}^{+} &= \langle  \chi_{\alpha}| V_{\rm tot} G_{0}(+\omega+i\eta+\varepsilon_{i})| \chi_{\beta} \rangle\\
  g_{i\alpha}[a,b] &= \langle  \chi_{\alpha}|f_{\rm HXC} \delta \rho[a,b]| \psi_{i}\rangle\\
  h_{i\alpha} &= \langle  \chi_{\alpha}| \delta V_{\mathcal P} |\psi_{i} \rangle
\end{align*}
The computation of $S$, $H$ and $h_{i}$ are standard. For $K$, the matrix elements are
not analytic. However, the values $(G_{0}(\pm\omega+i\eta+\varepsilon_{i}) \chi_{\beta})(\br)$ can be
computed analytically for all $\br$ (see Appendix). Therefore, we introduce an
integration grid with points $\br_{l}$ and weights $w_{l}$, and
approximate $K_{\alpha\beta}$ as
\begin{align*}
  K_{\alpha\beta} \approx \sum_{l} w_{l} \chi_{\alpha}(\br_{l}) V_{\rm tot}(\br_{l}) (G_{0}(\pm\omega+i\eta+\varepsilon_{i}) \chi_{\beta})(\br_{l}).
\end{align*}
The values of $V_{\rm tot}(\br_{l})$ arising from the nuclei and
exchange-correlation terms are computed exactly, as is usual in DFT.
Note that since the delocalizing operator $G_{0}$ is then multiplied
by localized quantities, the grid only needs to be sufficient to
integrated localized functions. In practice, we found that a coarse
exchange-correlation grid was often adequate.

The computation of $g$, assuming that all the $(a,b)$ are known, is
more conveniently reformulated as
\begin{align*}
  g_{i\alpha} = \langle  \delta \rho[a,b]|f_{\rm HXC} \psi_{i} \chi_{\alpha} \rangle.
\end{align*}
The values of $f_{\rm HXC} \psi_{i} \chi_{\alpha}$ on the grid are
precomputed, using the same technique as for the computation of $V_{\rm
  tot}$. Then, $\delta\rho[a,b](\br)  = 2 \sum_{j=1}^{N/2} \psi_{j}(\br)
(\delta\psi_{j}^{+}(\br) + \delta \psi_{j}^{-}(\br))$ is formed on the
grid, by using the values of $G_{0}(\pm\omega+i\eta+\varepsilon_{i})\chi_{\alpha}$ on the grid and the
$(a,b)$ coefficients.

The linear system
\eqref{eq:integral_eq_discretized}
is solved with the GMRES iterative solver,
preconditioned by $S-K_{i}^{+}$ (for the $a$ block) and
$(\pm \omega + \varepsilon_{i})S - H$ (for the $b$ block).

The additional computational cost compared to standard iterative TDDFT
computations is summarized in Table \ref{tab:cost}. Note that these steps
are cubic scaling (and, for low-lying excitations where the number of
ionized electrons is of the order of unity, quadratically scaling)
except for the steps involving $f_{\rm HXC}$. The scaling can be
improved using techniques such as the resolution of the identity.
However, in our tests, the step involving the computation of the
values of $G_{0} \chi_{\alpha}$ on the grid dominated the overall
computational time, and therefore we did not optimize the other steps.
\begin{table}[h!]
  \centering
  \begin{tabular}{|c|c|}
    \hline
    Operation&Cost\\
    \hline
    Precomputations of  $f_{\rm HXC} \psi_{i} \chi_{\alpha}$&$N_{\rm g} N_{\rm
                                                       occ} N_{\rm b} +
                                                       N_{\rm b}^{2}
                                                       N_{\rm g} N_{\rm
                                                occ}$\\
    Computation of $G_{0}(\pm\omega+i\eta+\varepsilon_{i}) \chi_{\alpha}$
    on the grid & $N_{\omega} N_{\rm ionized} N_{\rm g} N_{\rm b}$\\
    Matrix-vector products with $K$ & $N_{\omega} N_{\rm ionized} N_{\rm
                                      iter} N_{\rm g}
                                     N_{\rm b}$\\
    \hline
  \end{tabular}
  \caption{Dominant scaling of the main operations compared to
    standard iterative TDDFT. $N_{\rm b}$ is the number of basis
    functions, $N_{\rm g}$ the
    number of grid points, $N_{\rm occ}$ the number of occupied
    orbitals, $N_{\rm ionized}$ the number of ionized orbitals. Typically, $N_{\rm ionized} \ll N_{\rm occ} \ll N_{\rm b}
    \ll N_{\rm g}$. Additionally, $N_{\omega}$ is the
    number of frequencies desired, and $N_{\rm iter}$ is the number of
    iterations of the iterative solver (typically, $\le 10$).}
  \label{tab:cost}
\end{table}





\section{Results}
We implemented the method in the Julia programming language
\cite{bezanson2017julia}, interfacing with the PySCF package
\cite{sun2020recent} to perform the initial setup and DFT run. The integrals described in the
Appendix were implemented in the GaIn Fortran library. The code is
freely available at
\url{https://github.com/antoine-levitt/PhotoionizationGTO.jl} and
\url{https://gitlab.maisondelasimulation.fr/beDeft/GaIn}.

All results presented below use the LDA exchange-correlation
functional, with no spin polarization. The use of the LDA functional
is inadequate to obtain accurate photoionization spectra for the
systems studied here, but the emphasis in this paper is on the
methodology rather than on the particular results.

Unless explicitly mentioned, we used a very coarse
exchange-correlation grid (PySCF setting \texttt{1}), which yielded
reasonably accurate results at minimal computational cost. We used the
Dunning augmented basis sets \cite{dunning1989a, kendall1992a}, as
provided by the basis set exchange \cite{pritchard2019new}. These
basis sets are designed for converging post-Hartree-Fock methods
rather than TDDFT properties, and are very suboptimal here. We simply
use them to demonstrate that acceptable convergence can be obtained in
basis sets not specifically designed for that purpose.


\subsection{Atoms}
Atoms have an exponentially decaying total potential; therefore, we expect the $\delta\phi$ to be exponentially localized, making it an ideal case for our method. On atoms, we are able to compare the results to a reference (black line) computed using the atom-specific method of \cite{schwinn2022photoionization}. We focus in this section on \ch{He} and \ch{Be} atoms.

Helium is the simplest system, with only one occupied orbital (1s).
Accordingly, its TDLDA photoionization spectrum has only one
threshold, plotted as a dashed vertical line.
\begin{figure}[h!]
  \centering
  \includegraphics[width=.49\textwidth]{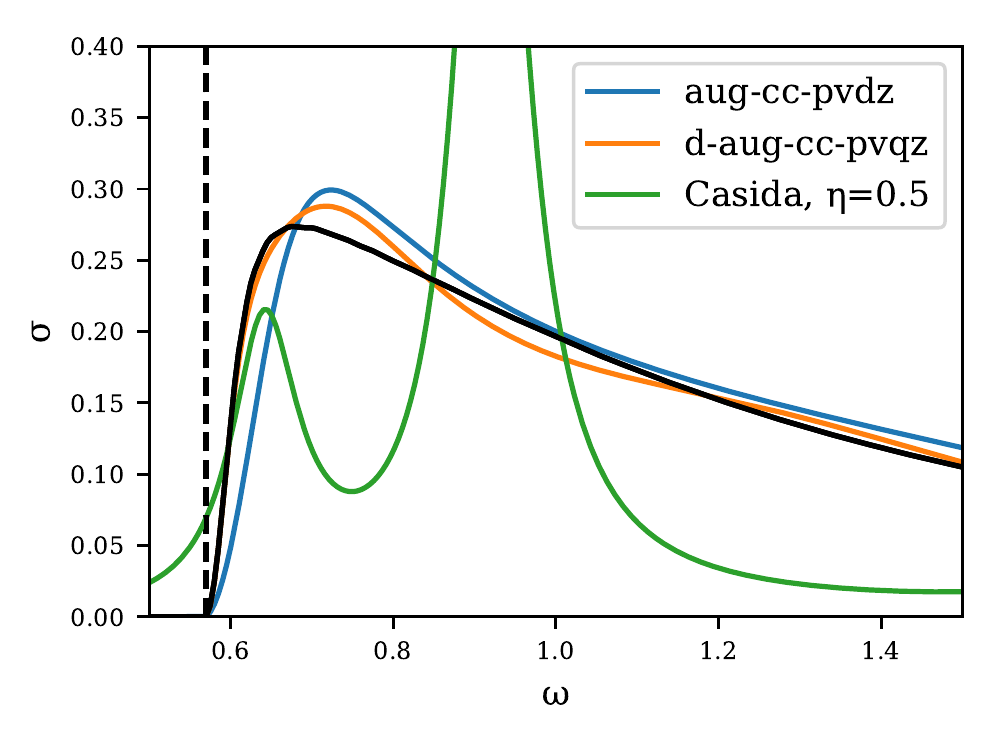}
  \includegraphics[width=.49\textwidth]{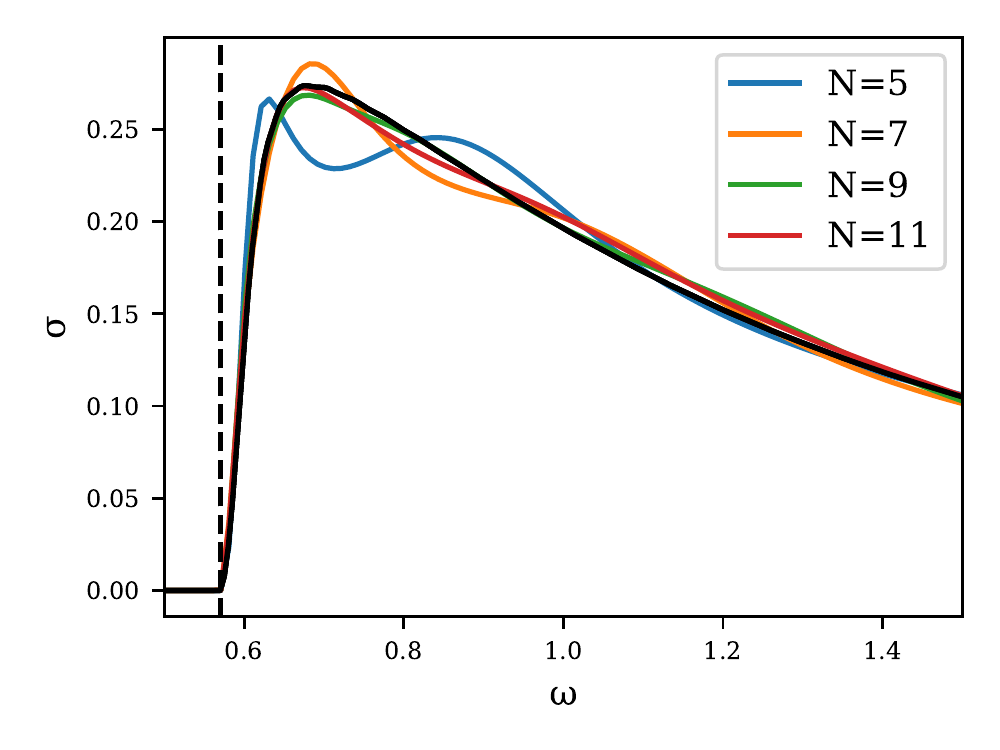}
  \caption{Photoionization of Helium, with standard basis sets (left)
    and even-tempered basis sets (right). The result of the standard
    Casida method with damping $(\eta=0.5)$ in the d-aug-cc-pvqz basis
    set is shown for comparison. Reference data (black line) from the
    method of \cite{schwinn2022photoionization}.}
  \label{fig:He}
\end{figure}
We see that results are already qualitatively consistent even at extremely
coarse basis sets: note for instance that the basis set aug-cc-pvdz
only has two basis functions for the p channel relevant here. To check
systematic convergence, we used a basis of even-tempered gaussians
with $N$ gaussian exponents logarithmically spaced from $0.01$ to $10$
(these values are taken somewhat arbitrarilly and not optimized).

To appreciate how inadequate the Casida method is to compute
photoionization spectra on this system, note that, in the largest
basis set used here, d-aug-cc-pvqz and in the frequency range
displayed here, there are only two relevant ($1$s $\to$ 2p)
excitations, at $\omega = 0.64$ and $\omega=0.91$. This is clearly not
sufficient to reconstruct a full spectrum. The same goes for other
approaches such as complex absorbing potentials; these methods, even with
optimal parameters, will only move these two poles in the
complex plane, which is not sufficient to reproduce the full structure of
the spectrum.

\begin{figure}[h!]
  \centering
  \includegraphics[width=.49\textwidth]{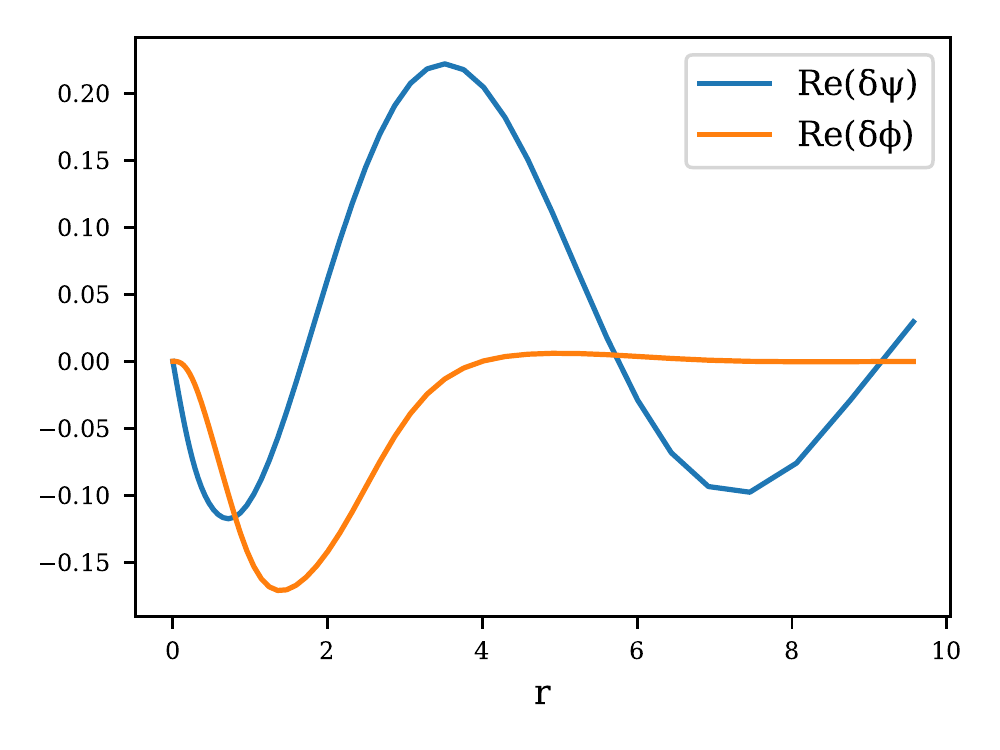}
  \caption{Orbital variations $\delta \psi^{+}$ and
    $\delta \phi^{+}$ (see text) for Helium at $\omega=1$.}
  \label{fig:He2}
\end{figure}

To illustrate, we plot Figure~\ref{fig:He2} for $\omega=1$ the orbital
variation $\delta \psi^{+}$, as well as the localized
$\delta \phi^{+}$ defined by
$\delta \psi^{+} = G_{0}(\omega+i0^{+}) \delta \phi^{+}$.
It is clearly not possible to represent the delocalized variation
$\delta \psi^{+}$ by localized orbitals, but $\delta \phi^{+}$ is
relatively short-range and therefore can reasonably be expanded on a gaussian basis set.

Beryllium has two occupied orbitals (1s and 2s), yielding two
different thresholds. Here, small standard basis sets are inaccurate, showing a displaced
peak after the $2s$ ionization and unphysical oscillations after the
$1s$ ionization. Using even-tempered basis set (using 10 gaussians
with exponents logarithmically spaced between 0.01 and 10) yields an
almost perfect ionization spectrum.

The ionization threshold of the 2s orbital lies below the 1s$\to$2p
excitation energy, which turns into a resonance. This resonance,
extremely hard to capture with standard damping methods, is present
and relatively accurate even with very coarse basis sets. Note that,
since we are able to compute analytic continuations of the matrix
elements of the free Green function as $z$ crosses the positive real
axis, we could compute this resonance directly
\cite{duchemin2023efficient}, but we do not pursue this direction in this paper.

\begin{figure}[h!]
  \centering
  \includegraphics[width=.49\textwidth]{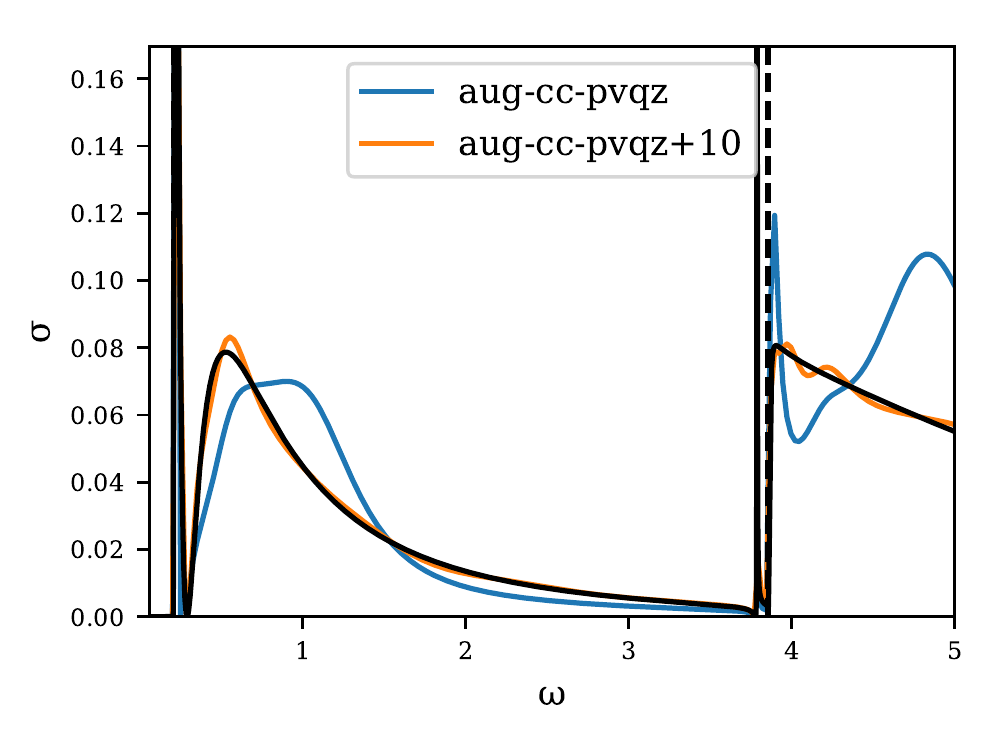}
  \includegraphics[width=.49\textwidth]{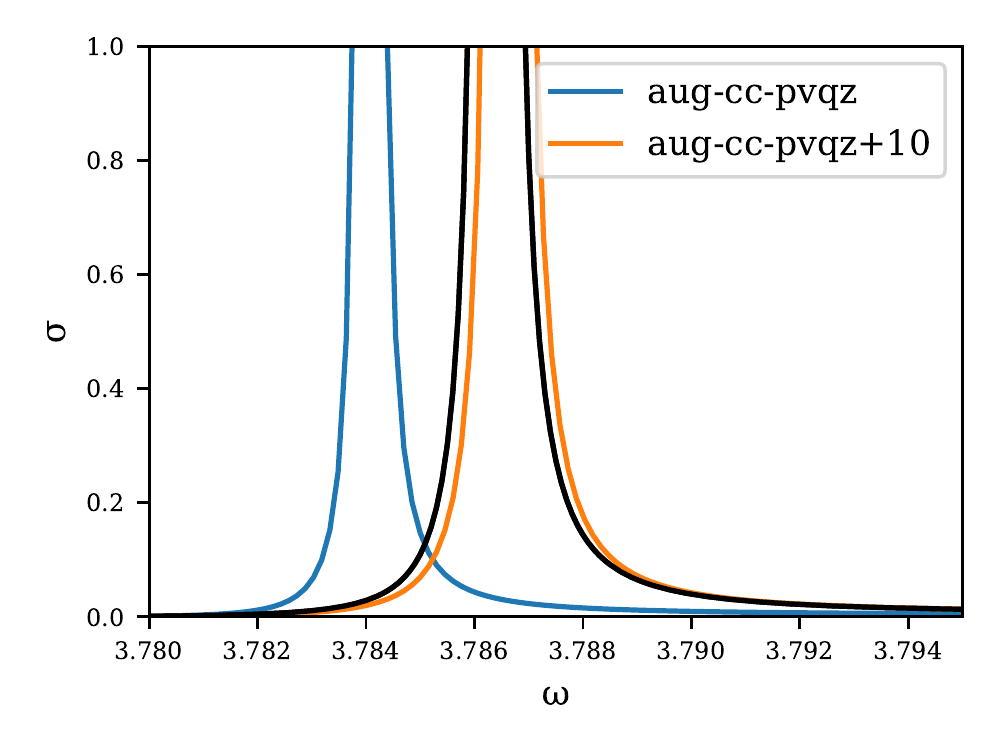}
  \caption{Photoionization of Beryllium (left), with zoom on the
    1s$\to$2p resonance (right).}
  \label{fig:Be}
\end{figure}



\subsection{Nonpolar molecules}

We next try our method on nonpolar molecules, on which the asymptotic
decay of the total potential is dictated by the quadrupole moment (decaying thus as
$1/r^{3}$). The test was conducted for \ch{H_2} and \ch{CH_4}, demonstrating a rapid convergence with respect to the basis set used: in both cases the spectrum is already
almost converged  at the aug-ccpvdz level. In particular, we find that moving from \emph{aug} to  
\emph{d-aug} basis sets turns out to be much more efficient, convergence-wise, than increasing the zeta label of the basis. This confirms again that the relevant components lies in the asymptotic part of the wave-function, supported by the delocalized atomic orbitals.  

\begin{figure}[h!]
  \centering
  \includegraphics[width=.49\textwidth]{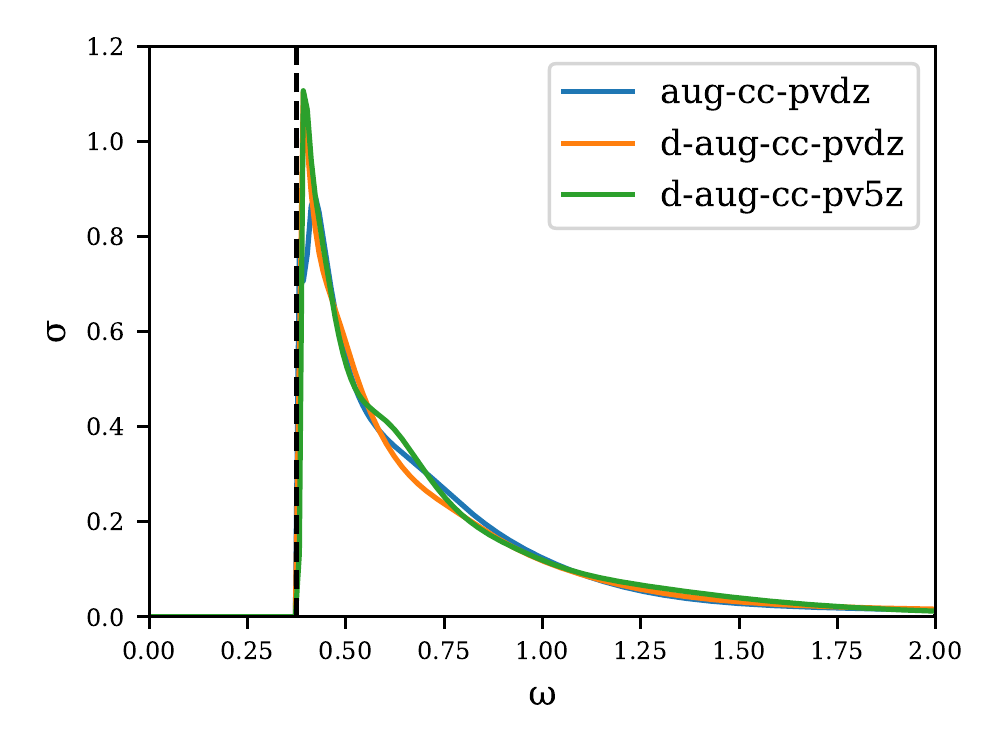}
  \caption{H2}
  \label{fig:H2}
\end{figure}

\begin{figure}[h!]
  \centering
  \includegraphics[width=.49\textwidth]{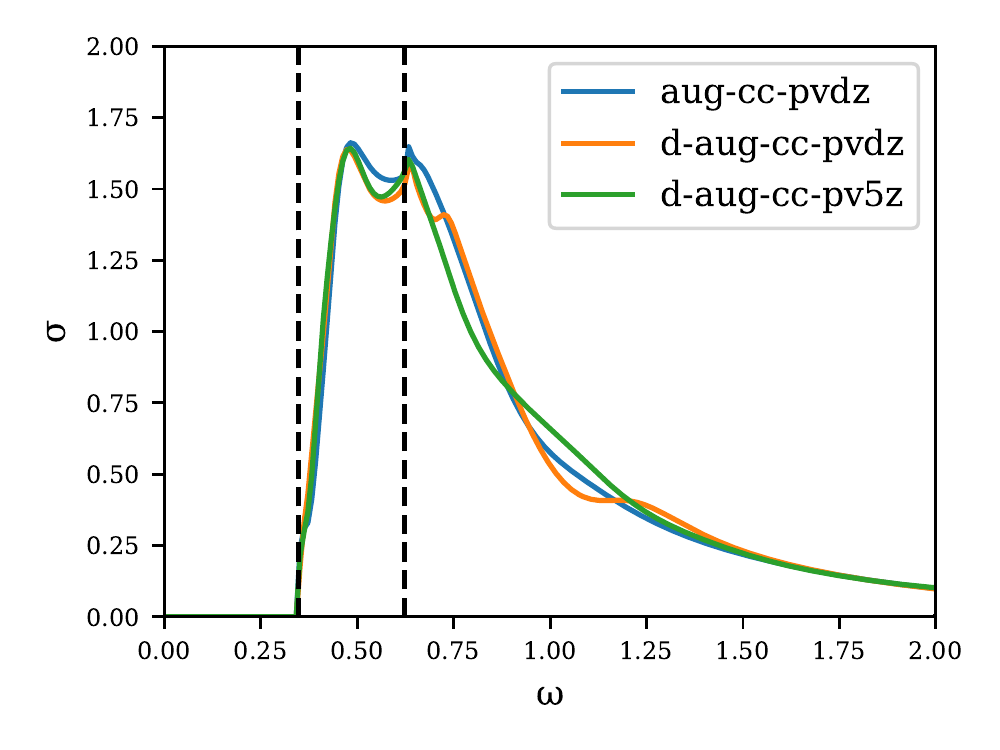}
  \caption{CH4}
  \label{fig:CH4}
\end{figure}


\subsection{Polar molecule}
We test the method on the strongly polarized \ch{LiH} molecule.
\begin{figure}[h!]
  \centering
  \includegraphics[width=.49\textwidth]{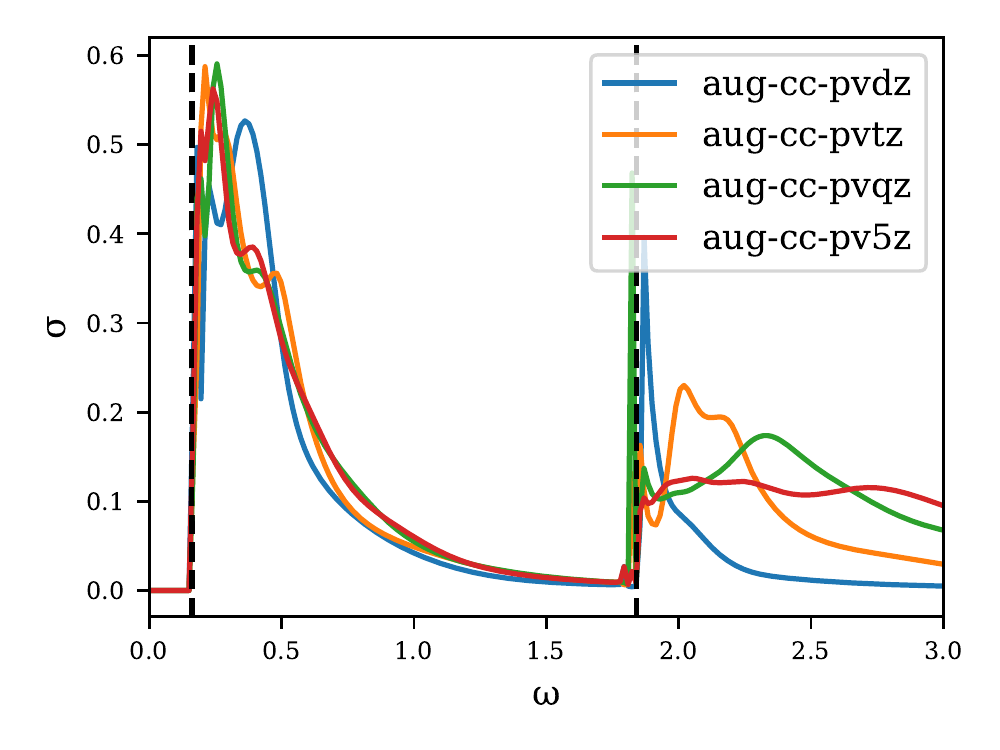}
  \caption{LiH}
  \label{fig:LiH}
\end{figure}
The strong dipole moment makes it hard for the method to converge,
especially after the second ionization. Spurious oscillations after
the peaks appear, which are consistent with what has been observed
using approximate boundary conditions in radial methods
\cite{schwinn2022photoionization}. However, we remark that these
difficulties do not hinder the capture of both ionization thresholds.

\subsection{Charged systems}

We use \ch{Li+} as a benchmark in this case. As any other charged
systems, ions have a long-range total potential decaying in $1/r$. In
such a case, the asymptotic form of the $\delta \psi$ is modified from
a plane wave to a Coulomb wave
\cite[Appendix]{schwinn2022photoionization}. Therefore, the functions
$\delta \phi$ become much more delocalized, and our method is inadequate.

\begin{figure}[h!]
  \centering
  \includegraphics[width=.49\textwidth]{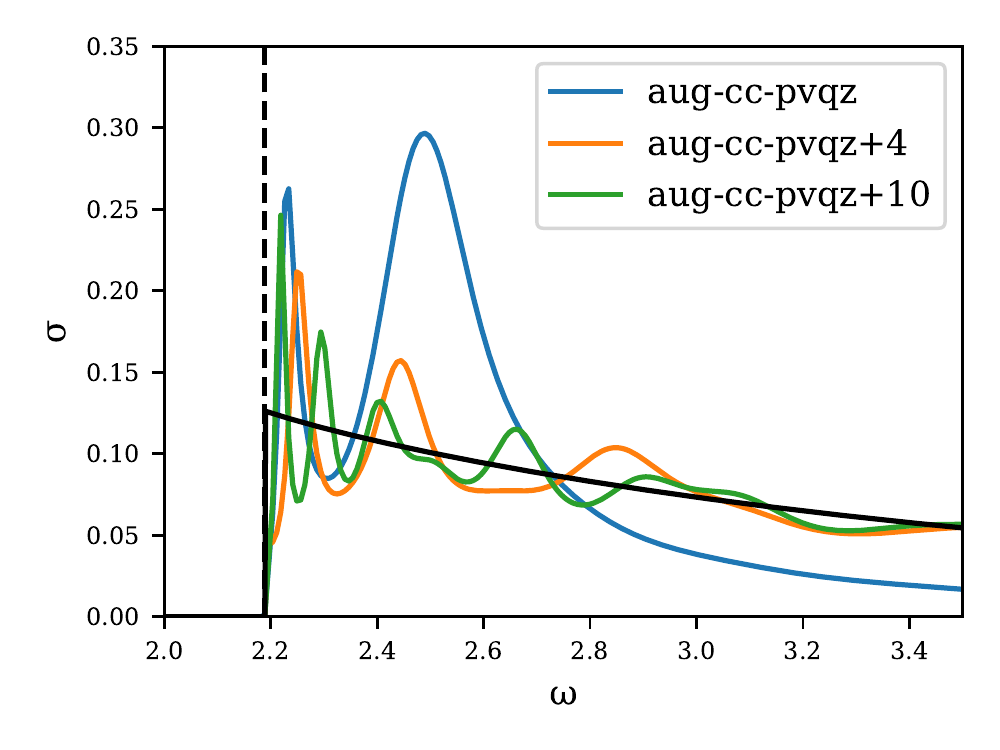}
  \caption{\ch{Li+}}
  \label{fig:Liplus}
\end{figure}
Accordingly, there are strong oscillations after the ionization
threshold, which disappear very slowly when increasing the basis set
size; this is consistent with Figure~5 of
\cite{schwinn2022photoionization}. For this example, the coarse
integration grids we used before were not sufficient to resolve the
fine oscillations, and we used a Gauss-Chebyshev grid with $100$
points in the radial direction produced by PySCF.


\section{Conclusion}
We have presented a new method to compute photoionization spectra for
TDDFT with semilocal exchange-correlation functionals in
gaussian basis sets, and tested it on atoms and small molecules. The
method appears to be very efficient on atoms and nonpolar molecules;
it struggles on polar molecules and charged systems.

We note the following possible improvements to our methodology
\begin{itemize}
\item For simplicity, we used standard Dunning basis sets, with exchange-correlation
  integration grids. Both of these were designed for a different
  problem (converging ground states orbitals and exchange-correlation
  integrals) than the one addressed here, and it is likely that
  computational efficiency can be significantly improved by tailoring
  these to our problem rather than using off-the-shelf technology.
\item The current methodology scales formally as the fourth power of
  the number of electrons, with the possibility of cubic scaling at
  the cost of a higher prefactor. It should be possible to reduce this
  scaling even further using techniques like the resolution of the
  identity.
\item Although we tested our scheme on semilocal density functionals,
  it is in theory straightforwardly adaptable to hybrid functionals.
  However, since the Sternheimer equation is effectively long-range in
  these cases \cite{schwinn2022photoionization}, we expect to face the
  same difficulties as we did in the ionic case. An efficient
  treatment of the Coulomb potential is, as far as we know, an open
  problem in the non-radial case. A partial solution could be to
  truncate artificially the potential; this would yield better basis
  set convergence and eliminate the oscillations, at the price of a
  regularization of the problem (particularly important near
  ionization thresholds).
\item We only explored linear response TDDFT in the frequency domain; it would be interesting
  to generalize this methodology to the non-perturbative TDDFT
  equations in the time domain. Recent progress has been made in this
  direction in \cite{kaye2022eliminating}, using a similar decoupling
  between the kinetic and potential operators.


\end{itemize}

Finally, we note that we have focused here on atoms and small
molecules. Experience suggests that it might be harder to converge
photoionization spectra of small molecules than of large systems,
because the relatively fine features computed here average out over a
molecule, and because the basis sets used for larger molecules cover a
larger region of space. Exploring this further is an interesting topic
for future research.

\section*{Acknowledgments}
We are grateful to Karno Schwinn and Julien Toulouse for the reference
data on atoms and ions.

\section*{Appendix: matrix elements of the Helmholtz kernel in
  Gaussian basis sets}
Our goal is the computation of integrals of the form
\begin{align*}
  I = \langle  g_{1}, G_{0}(\omega+i0^{+}) g_{2} \rangle = - \frac 1 {2\pi} \int_{\R^{6}} g_{1}(\br) g_{2}(\br') \frac{e^{-\lambda |\br-\br'|}}{|\br-\br'|}.
\end{align*}
when $g_{1}$ and $g_{2}$ are gaussian-type orbitals, for complex values
of $\lambda$. The case of interest in the present paper is
$\lambda = -i\sqrt{2\omega}$ (as well as possible analytic
continuations of this). This also includes pointwise values of
$G_{0}(\omega+i0^{+}) g_{2}$ by taking for $g_{1}$ the limit of a
zero-width gaussian. Explicit formulas have been obtained in the
context of range-separated hybrids with Yukawa potentials
\cite{ten2004initiation, ten2007new, akinaga2008range}; we recall here
the method of computation.

As is standard, it suffices to compute the integral for the primitive
radial gaussians $g_{1}(\br) = e^{-\alpha|\br-\bR_{1}|^{2}}$,
$g_{2}(\br) = e^{-\beta|\br-\bR_{2}|^{2}}$. Expressions for integrals
involving higher angular momenta can be obtained by differentiating
the integrals with respect to the centers $\bR_{1}$ and $\bR_{2}$.

We use the integral representation:
\[
\frac{e^{-\lambda x}}{x}=\sqrt{\frac{2}{\pi}}\int_0^\infty dt\, e^{-\frac{1}{2}(x^2t^2 +\frac{\lambda^2}{t^2})}
\]
and relatively straightforward but tedious changes of variables to obtain
\[
\begin{split}
I & = - \frac{1}{2} \left(\frac{\pi}{\alpha}\right)^{3/2}
\left(\frac{\pi}{\beta}\right)^{3/2} \frac{e^{ -a^2r^2 }}{r} \left[ \mathrm{erfcx}\left( \frac{\lambda}{2a}+ ar \right) 
- \mathrm{erfcx}\left( \frac{\lambda}{2a} - ar \right) \; \right]
\end{split}
\]
with $a=\sqrt{\alpha\beta/(\alpha+\beta)}$ the combined
Gaussian exponent, $r=|\bR_t-\bR_s|$ and where $\mathrm{erfcx}$ denotes the scaled complementary
error function: $\mathrm{erfcx(z)}=e^{z^2}\mathrm{erfc}(z)$.

The computation of derivatives requires special care for numerical
stability. We refer to \cite{ten2007new} for a discussion of the
issues and robust methods. We have implemented a similar method to
that of \cite{ten2007new} and generalized it to complex values of
$\lambda$. Since we have found the method of
\cite{molin:hal-00580855,al2021computation} to be very fast for the
computation of error functions, we found that some of the
complexity caused by the avoidance of these computations in
\cite{ten2007new} were not needed, and we use a slightly simplified
implementation, using an upward recurrence relation (s1 in
\cite{ten2007new}) for large values of $ar$, and a Taylor expansion in
$ar$ (s3 in \cite{ten2007new}) for small values.

\bibliographystyle{plain}
\bibliography{biblio.bib}
\end{document}